\documentclass{PoS}
\usepackage{xspace}
\newcommand{\Al}{$^{26}$Al\xspace}
\newcommand{\about}{$\simeq$}

\newcommand{\Msol}{M\ensuremath{_\odot}\xspace}

\let\jnl=\rmfamily
\def\refe@jnl#1{{\jnl#1}}%
\newcommand\aj{\refe@jnl{AJ}}%
\newcommand\actaa{\refe@jnl{Acta Astron.}}%
\newcommand\araa{\refe@jnl{ARA\&A}}%
\newcommand\apj{\refe@jnl{ApJ}}%
\newcommand\apjl{\refe@jnl{ApJ}}%
\newcommand\apjs{\refe@jnl{ApJS}}%
\newcommand\ao{\refe@jnl{Appl.~Opt.}}%
\newcommand\apss{\refe@jnl{Ap\&SS}}%
\newcommand\aap{\refe@jnl{A\&A}}%
\newcommand\aapr{\refe@jnl{A\&A~Rev.}}%
\newcommand\aaps{\refe@jnl{A\&AS}}%
\newcommand\azh{\refe@jnl{AZh}}%
\newcommand\memras{\refe@jnl{MmRAS}}%
\newcommand\mnras{\refe@jnl{MNRAS}}%
\newcommand\na{\refe@jnl{New A}}%
\newcommand\nar{\refe@jnl{New A Rev.}}%
\newcommand\pra{\refe@jnl{Phys.~Rev.~A}}%
\newcommand\prb{\refe@jnl{Phys.~Rev.~B}}%
\newcommand\prc{\refe@jnl{Phys.~Rev.~C}}%
\newcommand\prd{\refe@jnl{Phys.~Rev.~D}}%
\newcommand\pre{\refe@jnl{Phys.~Rev.~E}}%
\newcommand\prl{\refe@jnl{Phys.~Rev.~Lett.}}%
\newcommand\pasa{\refe@jnl{PASA}}%
\newcommand\pasp{\refe@jnl{PASP}}%
\newcommand\pasj{\refe@jnl{PASJ}}%
\newcommand\skytel{\refe@jnl{S\&T}}%
\newcommand\solphys{\refe@jnl{Sol.~Phys.}}%
\newcommand\sovast{\refe@jnl{Soviet~Ast.}}%
\newcommand\ssr{\refe@jnl{Space~Sci.~Rev.}}%
\newcommand\nat{\refe@jnl{Nature}}%
\newcommand\iaucirc{\refe@jnl{IAU~Circ.}}%
\newcommand\aplett{\refe@jnl{Astrophys.~Lett.}}%
\newcommand\apspr{\refe@jnl{Astrophys.~Space~Phys.~Res.}}%
\newcommand\nphysa{\refe@jnl{Nucl.~Phys.~A}}%
\newcommand\physrep{\refe@jnl{Phys.~Rep.}}%
\newcommand\procspie{\refe@jnl{Proc.~SPIE}}%
\title{Nucleosynthesis and Gamma-Ray Line Spectroscopy with INTEGRAL}

\ShortTitle{Nucleosynthesis}

\author{\speaker{Roland Diehl}\\% \thanks{A footnote may follow.}\\
        Max-Planck-Institut f\"ur extraterrestrische Physik, D-85741 Garching, Germany\\
        E-mail: \email{rod@mpe.mpg.de}}

\abstract{Cosmic nucleosynthesis co-produces unstable isotopes, which emit characteristic gamma-ray emission lines upon their radioactive decay that can be measured with SPI on INTEGRAL. 
High spectral resolution allows to derive velocity constraints on nucleosynthesis ejecta down to \about~100~km~s$^{-1}$. 
Supernova explosions generate gamma-rays from radioactive decays of $^{56}$Ni and $^{44}$Ti, which can reveal details of the explosion mechanism. 
Core-collapse supernovae apparently do not always produce significant amounts of $^{44}$Ti, as in the Galaxy fewer sources than expected from the supernova rate have been found. INTEGRAL's $^{44}$Ti data on the well-observed Cas A and SN1987A events are evidence that non-spherical explosions and $^{44}$Ti production may be correlated. 
INTEGRAL may have the chance to see directly for a SNIa the $^{56}$Ni decay chain which powers the supernova light curve: SN2011fe at only 6.4 Mpc distance may be close enough. 
Characteristic gamma-ray lines from radioactive decays of long-lived $^{26}$Al and $^{60}$Fe isotopes have been exploited to obtain information on the structure and dynamics of massive stars in their late evolution and supernovae, as their yields are sensitive to those details. 
The extended INTEGRAL mission establishes a database of sufficiently-deep observations of several specific regions of massive star groups, such as Cygnus, Carina, and Sco-Cen. In the inner Galaxy, $^{26}$Al nucleosynthesis gamma-rays help to unravel the Galaxy's structure and the role of a central bar, as the kinematically-shifted  $^{26}$Al gamma-ray line energy records the longitude-velocity behavior of hot interstellar gas. 
Thus, INTEGRAL has consolidated the feasibility of constraining cosmic nucleosynthesis through gamma-ray line observations.
High-resolution spectroscopy with SPI provides insights into supernova explosion physics, massive-star interiors, and dynamics of nucleosynthesis ejecta in extended interstellar space. Due to its extended mission INTEGRAL maintains its chance to also see rare sufficiently-nearby events; for example, a nearby nova could provide first nova nucleosynthesis measurements of $^7$Be and $^{22}$Na production.}

\FullConference{The Extreme and Variable High Energy Sky - extremesky2011,\\
		September 19-23, 2011\\
		Chia Laguna (Cagliari), Italy}
		
\begin{document}

\section{Nucleosynthesis and Gamma-Rays} %%%%%%%%%%%%%%%%%%%%%%%%%%%%%%%%%%%
Radioactive by-products of nuclear fusion reactions in astrophysical objects are the basis for one of INTEGRAL's key science objectives, the study of cosmic nucleosynthesis through gamma-ray line spectroscopy measurements. Candidate sources are supernovae, massive stars, and novae. Radioactive ejecta, in the case of short-lived ($\leq $My) radioactivities, decay in the exploding source during its expansion, or they produce an extended, diffuse emission for longer-lived radioactive isotopes and annihilating positrons. 
Early observations of the sky in gamma-rays had shown the existence of such lines, brightest lines being the positron annihilation line at 511 keV \cite{1975ApJ...201..593H}, %(Haymes et al. 1973), 
and the 1809~keV line from $^{26}$Al \cite{1982ApJ...262..742M}. % (Mahoney et al. 1982). 
Supernova 1987A in the LMC then led to first direct proof of supernova light origin from radioactive $^{56}$Ni decay \cite{1988Natur.331..416M}. % (Matz et al. 1988). 
This had prepared the ground for a first gamma-ray line sky survey with the Compton Gamma-Ray Observatory (CGRO; 1991-2000) with its OSSE and COMPTEL instruments. CGRO measurements first detected supernova radioactivity from $^{57}$Co in SN1987A \cite{1992ApJ...399L.137K}  and from $^{44}$Ti in Cas A \cite{1994A&A...284L...1I}. Diffuse radioactivity in $^{26}$Al was mapped over the Galaxy \cite{1995A&A...298..445D}, and positron emission gamma-rays were seen from the inner, bulge region of our Galaxy as an extended source \cite{1997ApJ...491..725P}. 
Yet, CGRO had a modest spectral resolution of order 10\%, inadequate for any gamma-ray spectroscopy in the sense of identifying new lines or constraining kinematics of source regions through line shape measurements; balloone-borne measurements had indicated the power of such measurements \cite{1990ApJ...351L..41T,1989Natur.339..122T} with solid state detectors operated at cryogenic temperatures. %The GRSE instrument originally foreseen as one of CGRO's detectors had been removed from the observatory as cost overruns made such trims necessary. 
INTEGRAL's spectrometer SPI was set out to perform such gamma-ray spectroscopy within the ESA Mission program. With a spectral resolution of 2--3~keV in the MeV regime, SPI is the highest spectral-resolution instrument ever operated for such observations, and will remain so for a while; no MeV-gamma-ray  spectroscopy mission is currently at the horizon of any of the international space agencies' programs. INTEGRAL data will establish the nuclear-astrophysics legacy database for (at least) the current generation of astrophysicists.

INTEGRAL's early-mission program included a core program part \cite{2003A&A...411L...1W}, where the plane of the Galaxy was surveyed, and regions of particular interest for nucleosynthesis studies were observed, such as the Cygnus and Vela regions, and the Cas A supernova remnant. The inner Galaxy itself obtained substantial exposure on top of this from monitoring programs on its rich population of transient X-ray sources. Altogether, at this time, INTEGRAL's sky exposure (Fig.~1) covers all regions of candidate sources of nucleosynthesis gamma-rays, although sensitivities in particular outside the inner-galaxy region are not always sufficient to constrain nucleosynthesis (see below).

%%%%%%%%%%%%%%%%%%%%%%%%%%%%%%%%%%%%%%%%%%%%%%%%%%%%%%%%%%%%%%%%%%%%%%%%%%%
\begin{figure}
\centering
\includegraphics[width=0.48\textwidth]{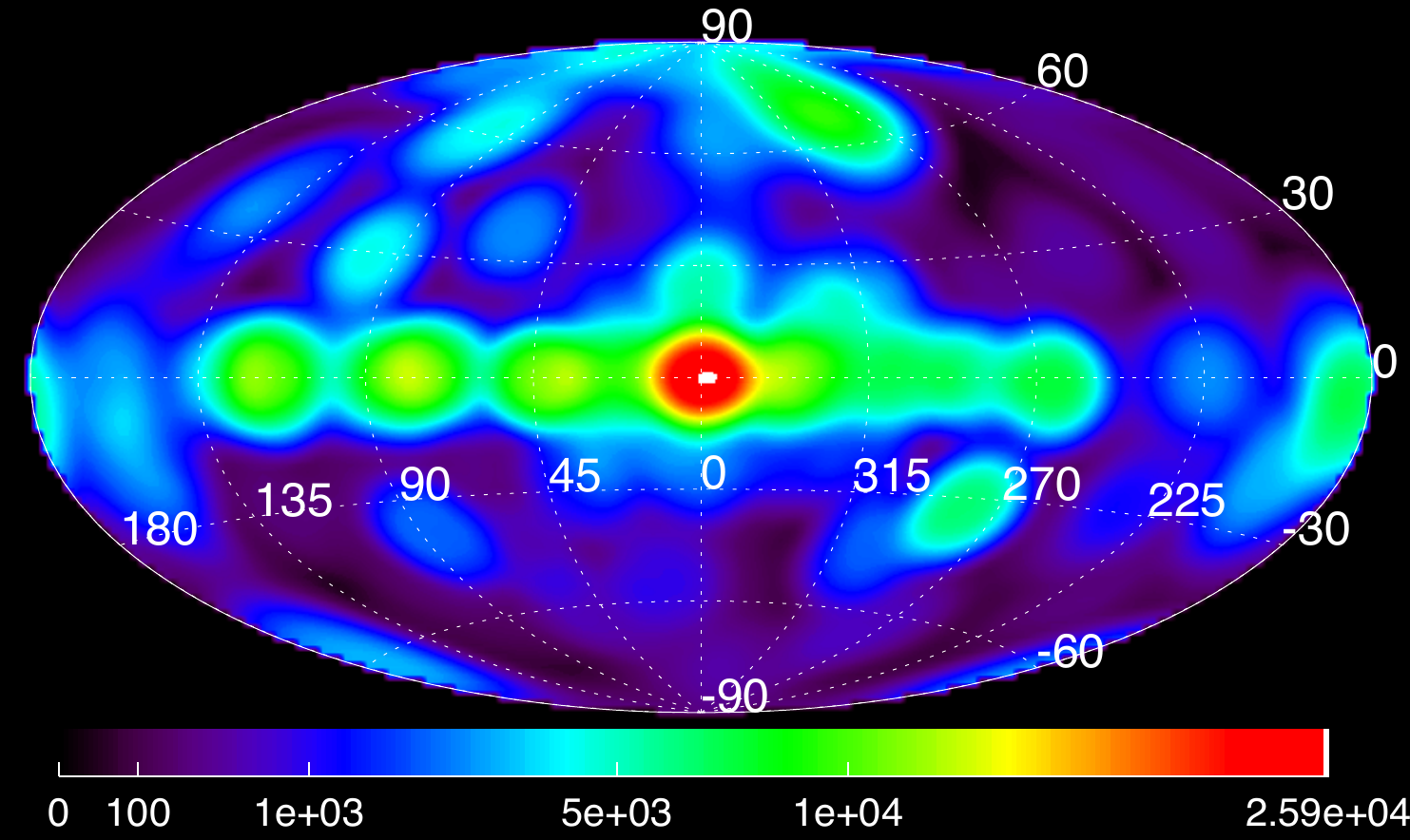}
\includegraphics[width=0.35\textwidth]{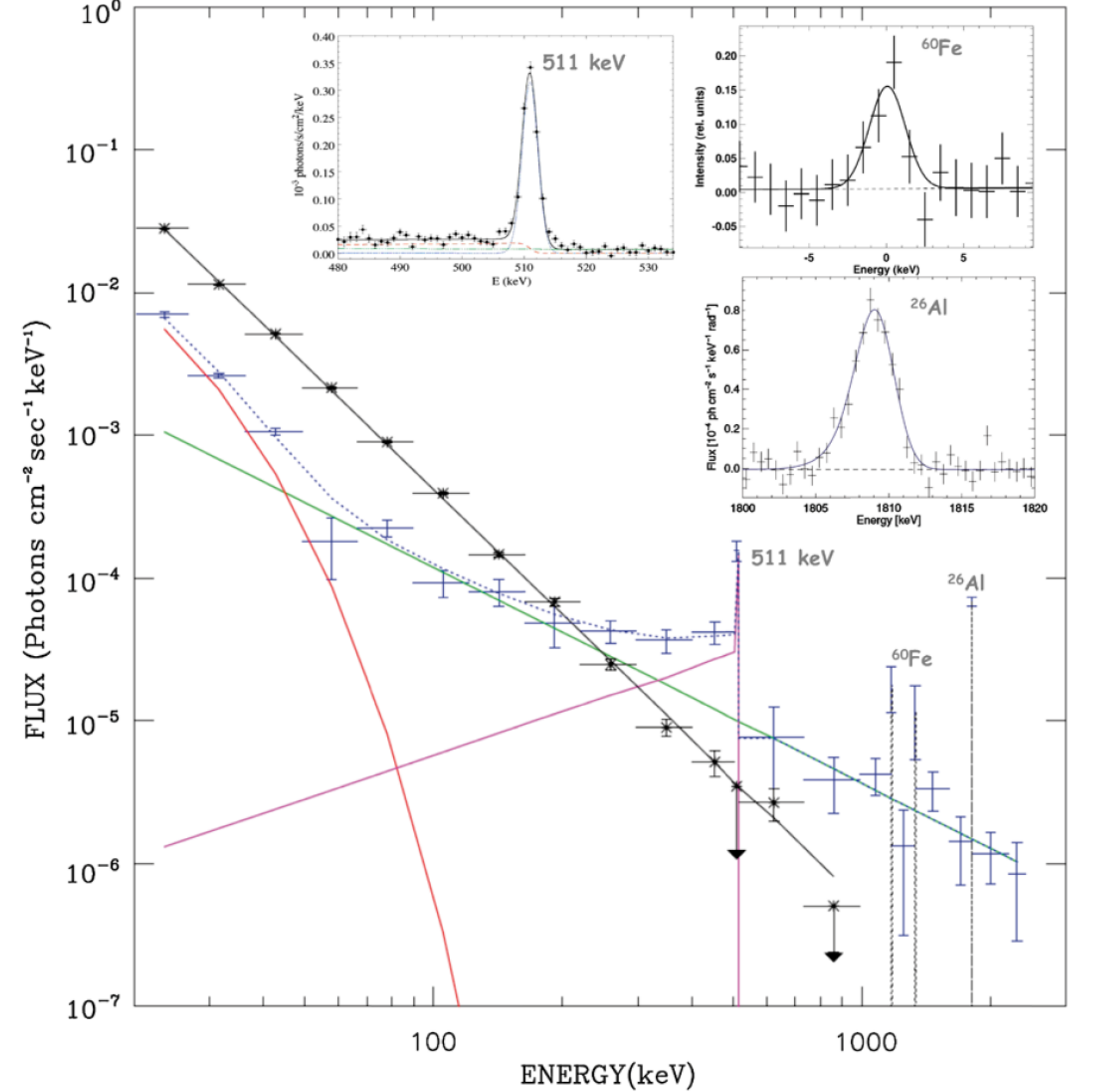}
\caption{The sky exposure of nucleosynthesis line observations with SPI after nine years of operations assembles data from the major candidate source regions, along the plane of the Galaxy, and the LMC and Virgo regions. \emph{(left)}. SPI measurements have observed diffuse-continuum and line emission from the Galaxy \cite{2011ApJ...739...29B,2011SSRv..161..149W}. \emph{(right)}.}
\label{fig:spi-exposure}
\end{figure}
%%%%%%%%%%%%%%%%%%%%%%%%%%%%%%%%%%%%%%%%%%%%%%%%%%%%%%%%%%%%%%%%%%%%%%%%%%%

\section{Supernovae} %%%%%%%%%%%%%%%%%%%%%%%%%%%%%%%%%%
We know of two basic types of supernova explosions, the thermonuclear explosion of a white dwarf star, which is believed to conform to observational supernova class SNIa \cite{2007Sci...315..825M}, and the core-collapse events, which are associated with all other observational classes of supernovae, SNIb-c and SNII with their subtypes \cite{2007PhR...442...38J}. Beyond such basic understanding, a physical explanation and model of neither of these supernovae types is established, although many candidate models are available. The problem lies in inadequate modeling of the observational variety, as these explosive events are highly dynamic and unfold into surrounding interstellar medium. Observational features are thus dominated by how the energy of the initial explosion and the radioactive decay of $^{56}$Ni therein is transferred into radiation of different types, as the photosphere of the object recedes and leads to transparency of the object on a time scale of months only \cite{2007MNRAS.375..154S}. Therefore, it remains difficult to relate observational features to the inner explosion physics. Neutrinos and gravitational waves are the basis of hopes to overcome this problem. Meanwhile, penetrating gamma-ray line and continuum emission appears to be most-directly related to physics in the supernova interiors, in particular as their brightness is driven by the weak interaction of radioactive decay and thus largely independent of thermal or density conditions and thus all expansion and ISM-interaction physics. This is the basis of their complementary value for supernova studies, and thus even few measurements in gamma-rays have a potential for major impacts on astrophysics of supernovae and their implications throughout astrophysical studies.

\subsection{Core-collapse Supernovae}
Massive stars end their evolution through gravitational collapse, if their initial mass exceeds about 8--10~\Msol \cite{2007PhR...442...38J,2011LNP...812..153T}. Radioactive ejecta which could be observed with gamma-ray spectrometers include $^{56}$Ni and $^{44}$Ti, at estimated amounts of \about~0.1 and \about~10$^{-4}$~\Msol, respectively. Although both isotopes should be produced in core collapse events which form neutron star remnants and hence eject freshly-produced isotopes, only the ejection of $^{56}$Ni appears secured as it powers the supernova light (although amounts vary by almost three order of magnitude \cite{2011Ap&SS.336..129N}), while $^{44}$Ti ejection is much less clear \cite{2006A&A...450.1037T}. INTEGRAL has surveyed the plane of our Galaxy \cite{2006NewAR..50..540R}, where one would expect to see about half a dozen $^{44}$Ti line sources if it would be ejected by typical core-collapse supernovae\cite{2006A&A...450.1037T}. None has been found beyond the 350-year old Cas A supernova remnant which had been discovered with COMPTEL. This relative paucity of $^{44}$Ti ejection suggests that core-collapse supernova explosions rarely occur with high spherical symmetry (as 1D models had to assume \cite{1996ApJ...464..332T}), but rather non-spherical with clumps and jets, where in particular $^{44}$Ti ejection would vary by more than an order of magnitude. \cite{2006A&A...450.1037T,2010ApJS..191...66M}.

%%%%%%%%%%%%%%%%%%%%%%%%%%%%%%%%%%%%%%%%%%%%%%%%%%%%%%%%%%%%%%%%%%%%%%%%%%%
\begin{figure}
\centering
\includegraphics[width=0.68\textwidth]{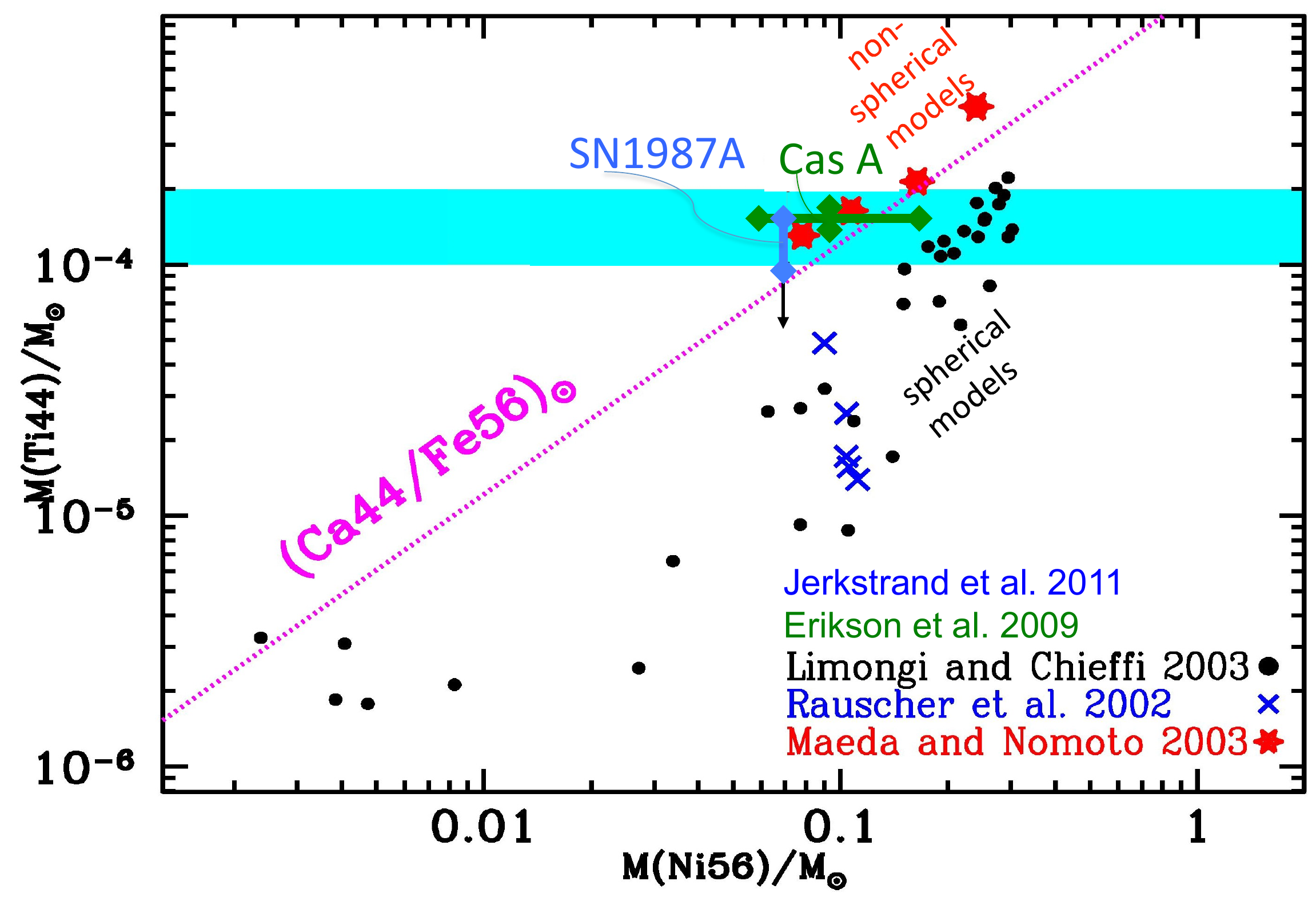}
\caption{The ratio of radioactive isotopes $^{56}$Ni and $^{44}$Ti is a sensitive probe of the effective location of the ``mass cut'', the separation between ejected material and the neutron-star remnant of a core-collapse supernova \cite{1996ApJ...464..332T}. Measurements for Cas A and SN1987A appear consistent with the ratio as inferred from solar abundances of the stable daughter isotopes. Models assuming spherical symmetry (black dots) fall below observations in this representation, while non-spherical explosion models indicate better agreement (adapted after \cite{2004ESASP.552...15P}).}
\label{fig:Ni-Ti-ratio}
\end{figure}
%%%%%%%%%%%%%%%%%%%%%%%%%%%%%%%%%%%%%%%%%%%%%%%%%%%%%%%%%%%%%%%%%%%%%%%%%%%

For the radioactivity afterglow of the Cas A supernova itself, the non-detection of the 1157~keV decay line by INTEGRAL's SPI instrument \cite{2009A&A...502..131M}, while INTEGRAL's IBIS measurement confirms an ejected $^{44}$Ti amount of 1-2~10$^{-4}$~\Msol \cite{2006ApJ...647L..41R}, suggests that $^{44}$Ti-rich ejecta move at velocities above 500~km~s$^{-1}$ \cite{2009A&A...502..131M}. As Doppler broadening increases with photon energy, this makes the high-energy decay line much broader than SPI's instrumental resolution, hence the signal-to-background ratio for this line degrades and may escape detection, while the 68 and 78~keV lines have been measured from this same event.  It will be interesting to see how Nustar \cite{2005ExA....20..131H} (to be launched in spring 2012) will map out the $^{44}$Ti emission morphology for Cas A, with its high imaging resolution, to complement these measurements and teach us more on supernova asymmetries in the case of Cas A. 

The detection of SN1987A in an energy band conforming to the 68 and 78~keV lines from $^{44}$Ti decay with IBIS was first reported at this conference (Grebenev et al., this volume). This provides an important second case, where for a well-observed supernova radioactivity in  $^{56}$Ni and $^{44}$Ti can be compared and confronted with model predictions for different assumptions of explosion symmetry (see Fig.~2; see also \cite{2011A&A...530A..45J,2002NewAR..46..487F,2010A&A...517A..51K}). 

\subsection{Supernovae of Type Ia}
SNIa are used as standard candles with known (and redshift-independent) brightness for cosmological studies \cite{2011ARNPS..61..251G}. This underlines the need for a physical model of this supernova type, at least to properly account for evolutionary biases such as could arise from progenitor's metallicity. Recently, the existence of substantial variety among SNIa has become established, and underlines the need for a physical understanding of explosion physics or at least of the physical origins of the Phillips relation used to standardize SNIa luminosities from their brightness evolution with time. Hope for an accurate measurement of the amount of $^{56}$Ni which powers a SNIa light curve (canonically about 0.7~\Msol, $\pm$0.3, \cite{2006A&A...450..241S}) through direct radioactive-decay gamma-rays had rested on the opportunity of a sufficently-nearby event, closer than about 5~Mpc \cite{2008NewAR..52..377I}. Those events should arise once every few years \cite{2008NewAR..52..377I}), depending on assumptions about the local galaxy populations. 

%%%%%%%%%%%%%%%%%%%%%%%%%%%%%%%%%%%%%%%%%%%%%%%%%%%%%%%%%%%%%%%%%%%%%%%%%%%
\begin{figure}
\centering
\includegraphics[width=0.68\textwidth]{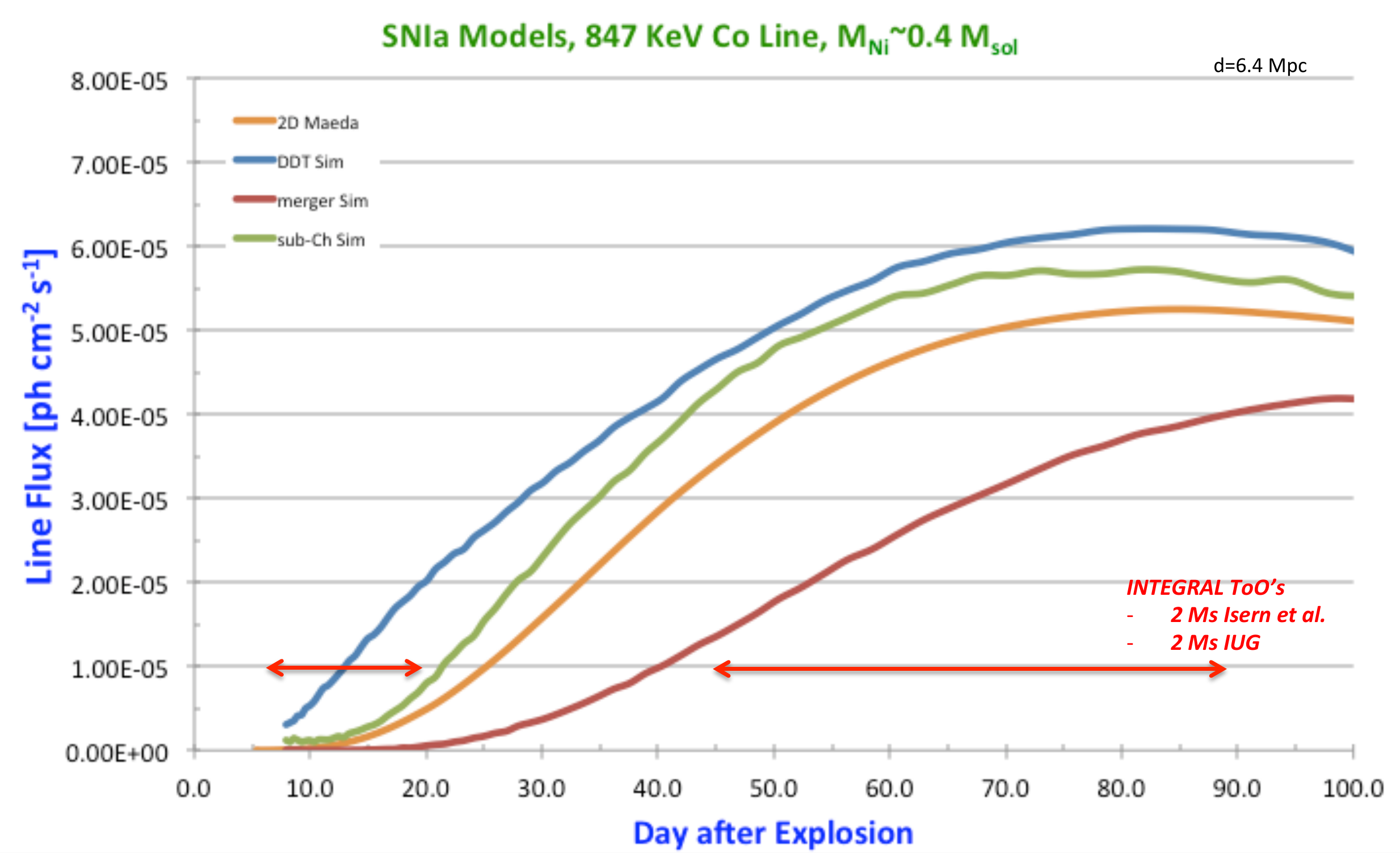}
\caption{The gamma-ray line emission at 847 keV from decay of $^{56}$Co rises towards a maximum at about 90 days after explosion. Shown here are the gamma-ray brightness evolutions according to several model variants for the case of SN2011fe (see legend for model types). The INTEGRAL observing windows are indicated by arrows in the lower part of the graph. INTEGRAL's narrow-line sensitivity would be 3~10$^{-5}$ph~cm$^{-2}$s$^{-1}$, degrading with line width to about 3~10$^{-4}$ph~cm$^{-2}$s$^{-1}$ if lines would be 50~keV wide.}
\label{fig:SN2011fe_847}
\end{figure}
%%%%%%%%%%%%%%%%%%%%%%%%%%%%%%%%%%%%%%%%%%%%%%%%%%%%%%%%%%%%%%%%%%%%%%%%%%%

So far,  only one SNIa, SN2003gs at 16~Mpc distance, had been considered a candidate for INTEGRAL observations, no other SNIa being closer and observable since then. This year's SN2011fe in M101/NGC5457 \cite{2011Natur.480..344N} appears a much more promising candidate, although still probably marginal with a distance of 6.4 Mpc, considering that lines are likely to be significantly broadened. But as nearby SNIa are rare, INTEGRAL observed this event for 4~Ms -- one Ms of observations was scheduled early after the explosion, in order to search for $^{56}$Ni which could be produced on the outer surface and appear early, if SN2011fe would be triggered by a He flash on a white dwarf's surface \cite{2010ApJ...714L..52S}. Then, a composite of 2+1~Ms were scheduled towards the expected maximum gamma-ray brightness, where one could hope for a line detection and possibly discriminate among candidate models \cite{2011PrPNP..66..309R} (see Fig.~3). Note that after the maximum, the SN is transparent to gamma-rays and explosion differences can be less constrained, as occultation differences vanish and only the total $^{56}$Ni mass determines the gamma-ray brightness. Analysis is currently underway, but first analyses do not show any of the candidate lines from the $^{56}$Ni decay chain \cite{2011ATel.3683....1I}.

\section{Diffuse Radioactivity from $^{26}$Al and $^{60}$Fe}
Massive-star interiors and supernovae are candidate sources of long-lived radioactive isotopes $^{26}$Al ($\tau\sim$~1.04~My) and $^{60}$Fe ($\tau\sim$~3.8~My) \cite{2011LNP...812..153T}. Both isotopes have been measured with SPI/INTEGRAL from the Galaxy at large \cite{2006A&A...449.1025D,2009A&A...496..713W,2007A&A...469.1005W} (Fig.~1), and $^{26}$Al is bright enough to also be seen from localized regions along the Galactic plane hosting many massive stars, such as Cygnus \cite{2009A&A...506..703M}. Nevertheless, the Cygnus region observations provide a key constraint on wind-ejected $^{26}$Al \cite{2009A&A...506..703M,2010A&A...511A..86M}.

Due to favorable location above the plane of the Galaxy, $^{26}$Al emission from the nearby (\about~120~pc) Scorpius-Centaurus association could be discriminated,  as discussed  in \cite{2010A&A...522A..51D}:
Stellar subgroups of different ages would result from a star forming region within a giant molecular cloud if the environmental effects of massive-star action of a first generation of stars (specifically shocks from winds and supernovae) would interact with nearby dense interstellar medium, in a scenario of  propagating or triggered star formation. Then later-generation ejecta would find the ISM pre-shaped by previous stellar generations. Such a scenario was proposed based on the different subgroups of the Scorpius-Centaurus Association \cite{1989A&A...216...44D, 2002AJ....124..404P} and the numerous stellar groups surrounding it \cite{2008A&A...480..735F}. 
Indications of recent star formation have been found in the L1688 cloud as part of the $\rho$~Oph molecular cloud, and may have been triggered by the winds and supernovae causing the \Al we observe. The young  $\rho$~Oph stars then could be interpreted as the latest signs of propagating star formation originally initiated from the oldest Sco-Cen subgroup in Upper Centaurus Lupus \cite{2008hsf2.book..351W}.  Many proposed scenarios of triggered star formation are only based on relatively weak evidence, such as the presence of Young Stellar Objects (YSOs) near shocks caused by massive stars.  
Positional evidence alone is not unequivocally considered to prove a triggered star formation scenario. Much more reliable conclusions can be drawn if the ages of the young stellar populations can be determined and compared to the moment in time at which an external shock from another star formation site arrived. Agreement of these timings would add convincing evidence for the triggered star formation scenario. 

 $^{60}$Fe is only marginally seen from the Galaxy as a whole, its brightness being only 1/7~ of that of $^{26}$Al \cite{2009A&A...496..713W,2007A&A...469.1005W}. In steady state, this brightness ratio constrains massive-star interiors globally, as those same sources produce each of these isotopes in different inner regions and at different times of stellar evolution -- yields from models for those object types should come out consistent for all their nucleosynthesis products \cite{2011LNP...812..345D}. Particularly interesting would be the $^{60}$Fe/$^{26}$Al ratio for source populations of specific ages, as the ratio varies significantly due to wind-released $^{26}$Al before any core-collapse supernova would eject $^{60}$Fe and more $^{26}$Al \cite{2009A&A...504..531V}. $^{60}$Fe is exclusively released in supernovae, although predominantly produced in the late shell-burning phase before the collapse of the core. Only the Cygnus region appears within INTEGRAL's sensitivity range for this, however \cite{2010A&A...511A..86M}. 

%%%%%%%%%%%%%%%%%%%%%%%%%%%%%%%%%%%%%%%%%%%%%%%%%%%%%%%%%%%%%%%%%%%%%%%%%%%
\begin{figure}
\centering
\includegraphics[width=0.48\textwidth]{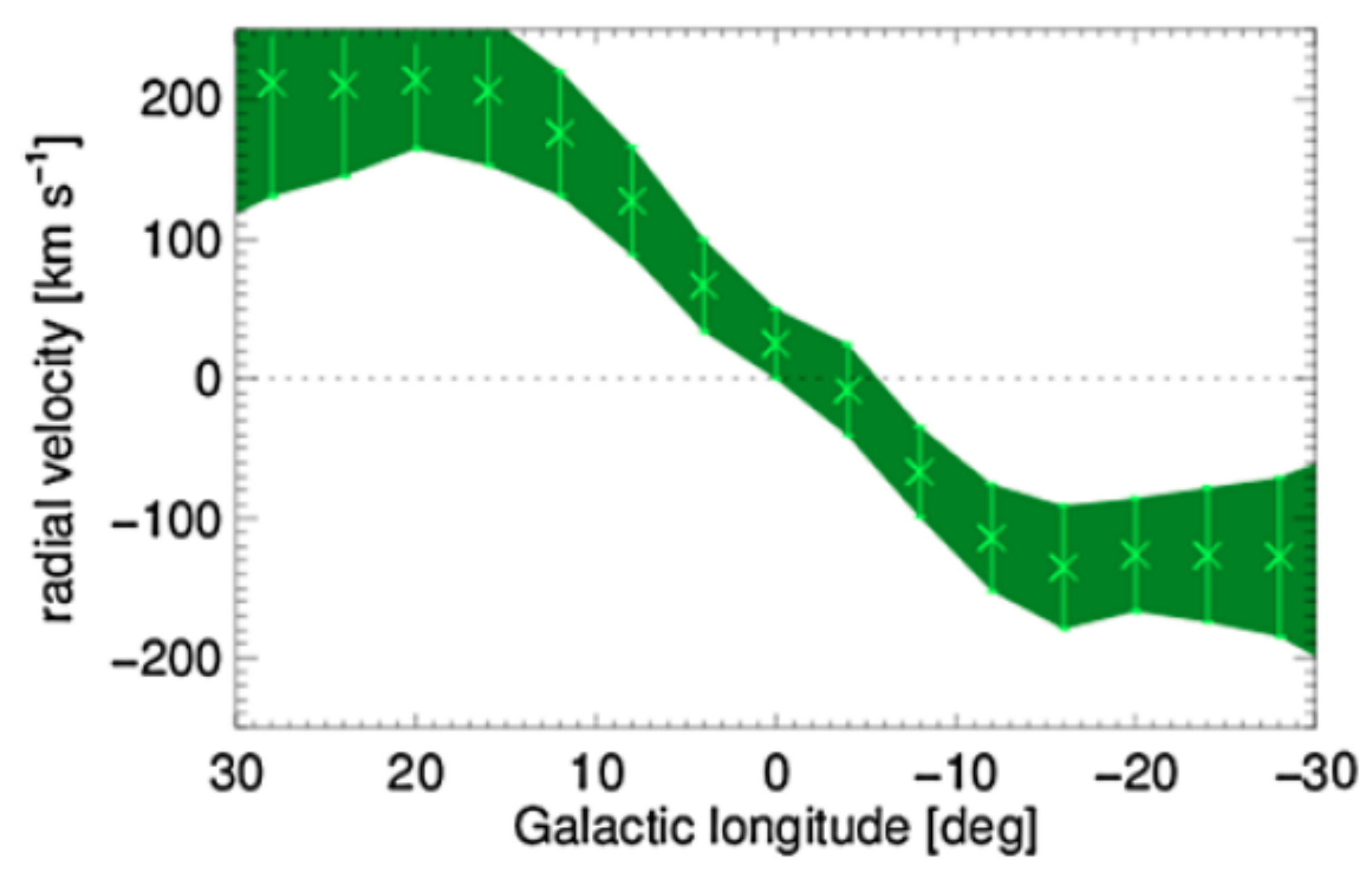}
\caption{The centroid of the $^{26}$Al line appears systematically Doppler-shifted along the inner Galaxy, as plausible from large-scale Galactic rotation. Note that data points as shown are not independent, but chosen to include sufficient $^{26}$Al signal and to maximize the longitudinal signature of $^{26}$Al line shifts (Figure from \cite{2011SSRv..161..149W}1).}
\label{fig:26Al_long-vel}
\end{figure}
%%%%%%%%%%%%%%%%%%%%%%%%%%%%%%%%%%%%%%%%%%%%%%%%%%%%%%%%%%%%%%%%%%%%%%%%%%%

%\subsection{Inner Galaxy}
Along the inner Galaxy, the uniquely-high spectral resolution of SPI allows an interesting application: Kinematic shift of the $^{26}$Al line from the Doppler effect and due to large-scale rotation of sources about the center of the Galaxy has been observed \cite{2006Natur.439...45D}.  The velocity resolution obtainable from $^{26}$Al gamma-rays now reaches 100~km~s$^{-1}$, which is astrophysical significant for tracing Galactic gas streams, or potentially constraining turbulence within supernova-blown  cavities of the interstellar medium. In the inner Galaxy, $^{26}$Al measurements (Fig.~4) record higher velocities at large scale than was derived from CO measurements for cold, molecular gas. This appears surprising, but massive star clusters along the bar of our Galaxy could plausibly lead to such kinematic behavior (Kretschmer et al., in preparation).

\section{Summary and Prospects} %%%%%%%%%%%%%%%%%%%%%%%%%%%%%%%%%%%

Observing ejection of freshly-produced radioactive isotopes still remains the major target of INTEGRAL, should the opportunity arise of a sufficiently nearby supernova or nova. The gamma-ray lines expected from current nucleosynthesis have been detected with INTEGRAL. For $^{26}$Al, now also studies of nucleosynthesis in identified massive-star groups (with therefore known source numbers, distances, and ages) became possible with INTEGRAL. $^{60}$Fe has first been measured by INTEGRAL with astrophysically-significant precision. $^{44}$Ti has now been seen from two core-collapse supernovae, and constrains nuclear yields and non-sphericities in inner regions of such supernovae. INTEGRAL so far did not have a chance to demonstrate the potential of gamma-ray measurements of supernova-powering $^{56}$Ni decay, due to paucity of sufficiently-nearby events (\about~5~Mpc for SNIa). Similarly, nova nucleosynthesis could not be constrained through $^{22}$Na nor $^{7}$Be, though positron annihilation provides some clue \cite{2000MNRAS.319..350J,2005A&A...441..513K}. Meanwhile, the deepening of exposure resulting from the extended mission now gradually lifts details beyond thresholds of scientific significance. Astrophysical conclusions are more precise than global galactic averages, and SPI's imaging capability is essential here. Continued deepening of exposure with INTEGRAL in its late mission years in such key regions will help to consolidate what could be learned from INTEGRAL on cosmic nucleosynthesis. 

{\bf Acknowledgements}
The INTEGRAL project of ESA and the SPI project has been completed under the responsibility and leadership of CNES/France.We are grateful to ASI, CEA, CNES, DLR, ESA, INTA, NASA and OSTC for support.
      The SPI anticoincidence system is supported by the German government through DLR grant 50.0G.9503.0. 
      %Analysis work for this paper was supported by the Munich Cluster of Excellence on ``Origins and Evolution of the Universe''.

%%%%%%%%%%%%%%%%%%%%%%%%%%%%%%%%%%%%%%%%%%%%%%%%%%%%%%%%%%%%%%%%%%%%%%%%%%%%%%%%%%%%%%%%%%%%%%%%%%%%%%%%%%%%%%%%%%%%%%%%%%%%
%
%\bibliographystyle{JHEP_MDL}
\bibliographystyle{plain}
%

% \bibliography{rod-references}

\end{document}